\documentclass[prl,twocolumn,showpacs,preprintnumbers,amsmath,amssymb]{revtex4}

\usepackage{graphicx}
\usepackage{dcolumn}
\usepackage{bm}

\begin{document}

\preprint{}

\title{Density-Dependent Response of an Ultracold Plasma to Few-Cycle Radio-Frequency Pulses}

\author{Truman M. Wilson, Wei-Ting Chen and Jacob L. Roberts}
 
\affiliation{%
Department of Physics, Colorado State University, Fort Collins, CO 80523
}%

\date{\today}

\begin{abstract}
Ultracold neutral plasmas exhibit a density-dependent resonant response to applied radio-frequency (RF) fields in the frequency range of several MHz to hundreds of MHz for achievable densities.  We have conducted measurements where short bursts of RF were applied to these plasmas, with pulse durations as short as two cycles. We still observed a density-dependent resonant response to these short pulses, but the timescale of the response is too short to be consistent with local heating of electrons in the plasma from collisions under a range of experimental parameters. Instead, our results are consistent with rapid energy transfer from collective motion of the entire electron cloud to individual electrons.  This collective motion was also observed by applying two sharp electric field pulses separated in time to the plasma.  These measurements demonstrate the importance of collective motion in the energy transport in these systems.
\end{abstract}

\maketitle
The creation of ultracold plasmas (UCPs) \cite{Killian1999} from photo-ionized, laser-cooled atoms has provided a way to study the dynamic processes of a freely expanding plasmas at cold temperatures.   One class of responses, collective oscillations, are a fundamental feature of plasma systems and can play a crucial role in energy transport as well as determining the response of a plasma to an external perturbation.  Collective oscillations were among the first reported experimental measurements from UCPs \cite{Kulin2000}, and have been subsequently observed in oscillatory behavior of UCP electron escape signals as the UCP evolves \cite{Castro2010,Zhang2008,Saleem2008,Chen2004,Mendonca2008,Shukla2010}.  They have also been excited using RF fields \cite{Kulin2000,Fletcher2006,Bergeson2003}, allowing for the UCP expansion rate to be measured \cite{Kulin2000,Zhang2008-2}.  From the expansion rate, it is possible to infer the early-time electron temperature \cite{Gupta2007}. 

The reason the UCP expansion rate can be measured through the application of an RF field is because plasma oscillations are density-dependent.  For an infinite, uniform density plasma the resonance condition for cold electron temperatures is given by $\omega_p = \sqrt{e^2n_e/m_e\epsilon_0}$, where $e$ is the elementary charge, $n_e$ is the local charge density, $\epsilon_0$ is the permittivity of free space, and $m_e$ is the mass of an electron.  However, UCPs do not have uniform densities, so this resonant frequency condition cannot be applied directly.  The relationship between the non-uniform plasma density and its resonant response to RF fields has been the subject of experimental \cite{Twedt2012} and theoretical work \cite{Bergeson2003,Lyubonko2012}.

In such previous work, the RF fields were assumed to be applied continuously throughout the UCP expansion.  In contrast, we apply short bursts of RF with as few as 2 cycles to the UCP in this work.  We still observe a density-dependent resonant response to to these short RF pulses.  Because the pulse length is short, we can easily observe the delay between the application of the RF and the subsequent electron escape signal.  This delay is typically $\sim$250 ns from the initial application of the RF to the peak electron escape signal, and does not vary significantly with electron temperature or UCP density.  These observations led us to develop a model of the UCP response to short-cycle RF pulses in which a collective motion of the entire electron cloud is the main mechanism for energy transfer to the escaping electrons in the UCP.  This collective motion of electrons induces electric fields with sufficient magnitude and spatial variation to cause electrons to escape the UCP.  These fields cannot be effectively screened by the plasma electrons because the oscillation frequency is by definition on the order of $\omega_p$.  This energy transfer mechanism is different than the ones assumed in \cite{Twedt2012,Bergeson2003,Lyubonko2012}.  This mechanism is prevalent in our system because we have UCP densities which are 1-2 orders of magnitude lower than in other experiments \cite{Killian1999,Bergeson2011,Simien2004}.  In this article, we describe observations of the UCP response to short bursts of RF and the model we developed for them.  Additional observations of the UCP response to pairs of sharp electric field pulses are also described and are consistent with the model that we have developed for the UCP response.

Before discussing the UCP response to a short burst of RF, it is useful to review the response of the UCP to a continuous application of the RF. One way this response is observed is through an increase in the electron escape rate from the UCP \cite{Kulin2000}.  As the UCP expands, the density drops putting some part of the density distribution at the resonance condition.  In previous experimental and theoretical work \cite{Kulin2000,Lyubonko2012,Twedt2012}, it is implied that the UCP electrons gain energy from the resulting oscillations through ohmic heating (i.e. collisional damping via electron-ion collisions).  This heat is then presumably collisionally redistributed \cite{Smorenburg2012}, promoting some electrons to energies that can overcome the electrostatic space charge, provided by the excess of ions \cite{Killian1999}, allowing them to escape.

Such collisional redistribution of the energy does not occur instantaneously, but rather on timescales associated with the electron self-equilibration time, which scales with both the density and temperature of the plasma \cite{Spitzer1962}.  These timescales can be calculated for the electron temperatures and densities that can be achieved with our apparatus, and range from 10s of ns to several $\mu$s.  Additionally, since the escaping electrons have a higher energy than the average electron energy, it is expected that the timescales associated with energy transfer to these electrons will be longer than this self-equilibration time \cite{Robicheaux2003}, as illustrated by evaporation rates in a gas of trapped anti-protons \cite{Andresen2010}.  Our observations of the UCP response timescales to short bursts of RF indicate that they do not vary under a wide range of UCP density and temperature conditions.  For a wide range of experimental parameters, the response is significantly faster than the expected equilibration timescales.

The UCPs for our work were created from the photoionization of ultracold $^{85}$Rb atoms.  The experimental sequence consisted of loading the Rb atoms in a magneto-optical trap (MOT) \cite{Chu1998}, then transferring them in a magnetic trap to a separate part of the vacuum system to be ionized.  The MOT was created using standard techniques \cite{Raab1987}.  From the MOT, the atoms were loaded into a magnetic quadrupole trap mounted on a translation stage.  The magnetically trapped atoms were then transferred $\sim$1 m \cite{Lewandowski2003} to a region in the vacuum system where electrodes can produce sensitive electric fields.
	
These cylindrically symmetric electrodes (Fig. 1) can be used to produce a variety of electric field configurations to extract escaping electrons from the UCP as well as to produce RF fields to induce plasma oscillations \cite{Fletcher2006,Bergeson2003,Kulin2000,Zhang2008-2}.  In this region, we also have the ability to apply magnetic fields both transversely and axially with respect to the electrodes.  It is important to note that the various grounded surfaces will attenuate applied potentials from the off-center electrodes at the location of the plasma.  An applied potential of 1 V at point A in Fig. 1 is less than 50 mV at point B.  The electric field from this configuration is reduced by a factor of 6 from a simple $E$ = $V$/$d$ calculation.

During the experiment sequence, the magnetic trap was turned off and the Rb atoms were ionized in a two-step photoionization process involving a resonant 780-nm laser and a pulsed dye laser at wavelengths of 471-479 nm ($\Delta$E/k$_B$ = 10-500 K above threshold) that controlled the initial electron energy.  We start with an approximately spherical Gaussian ion/electron density distribution that we describe by $n(r)$ = $n_0exp(-r^2/2\sigma^2)$, where $n_0$ is the peak density, and $\sigma$ is the rms radius.  In our system, we have initial peak plasma densities of $n_i$ = $n_e$ = 10$^7$ to 10$^8$ /cm$^{3}$ and an rms radius of $\sim$1 mm.  The density is 1-2 orders of magnitude lower than in other UCP experiments \cite{Killian1999,Bergeson2011,Simien2004}, and can have electron-electron self equilibration times that are more than a $\mu$s for sufficiently high electron temperatures and low enough densities.  For our achievable experimental conditions, we have plasma lifetimes that can be as long as $\sim$100 $\mu$s for our lowest initial ionization energies.  Upon ionization, a fraction of the electrons escaped creating a potential well which traps the remaining electrons, forming a UCP \cite{Killian1999}.  We define the fraction of electrons that have escaped prior to any point in the plasma evolution as the charge imbalance, $\delta$ = $(N_i-N_e)/N_i$, where $N_i$ and $N_e$ are the total number of ions and electrons in the UCP respectively.  Using the electrodes, we applied an electric field ($\sim$1 V/m) which pulled escaping electrons toward a microchannel plate detector (MCP) with the help of a mild guiding magnetic field ($\sim$7 G), which is axially symmetric with our electrode assembly \cite{Zhang2008-2}.  While this magnetic field helped guide electrons to our detector, we periodically made comparisons between data taken with and without the magnetic field to make sure none of the results presented in this paper were linked solely to the magnetic field.  In all cases, the UCP exhibited the same general behavior with and without the magnetic field present.  The MCP ultimately resulted in a current across a load resistance.  This produces a voltage that was calibrated to a number of electrons reaching the detector by measuring the value of the threshold voltage for the initial trapping of electrons as in \cite{Killian1999} over a range of initial ionization energies.  Extraction of the electrons to the MCP results in a time-of-flight delay, which we measured and took into account for all of the relevant times presented in this work.  

\begin{figure}
\includegraphics{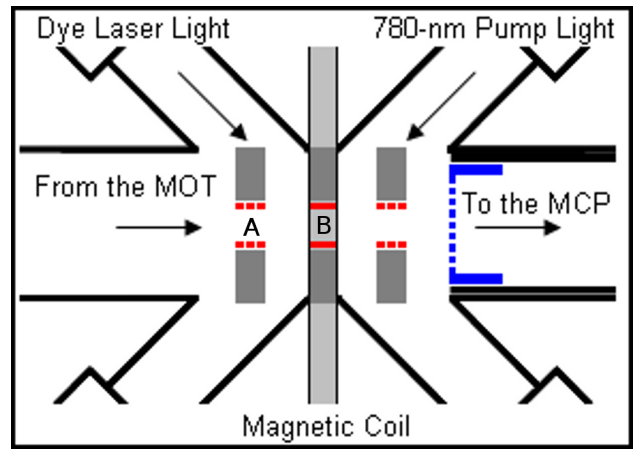}
\caption{(Color Online) A diagram of our electrode assembly.  The small red rectangles represent a cross section of the copper rings that we use for our electrodes.  The grey rectangles around them are the aluminum mounts, which are electrically grounded.  The magnetically trapped ultracold atoms are transferred to the center of these electrodes and ionized.  The electrodes and a set of wire mesh grids (blue) apply electric fields which pull electrons toward the MCP.  The distance between points A and B in the figure is 1.5 cm.}
\end{figure}
	
During the UCP expansion, we could apply an RF field either continuously throughout the expansion of the UCP or in a burst at a specific point in the plasma evolution.  The response of the UCP was measured via the escape of additional electrons as compared to the case without the RF applied.  The signals were collected on a fast oscilloscope and the peak response and integrated signals were analyzed to characterize the UCP response.

\begin{figure}
\includegraphics{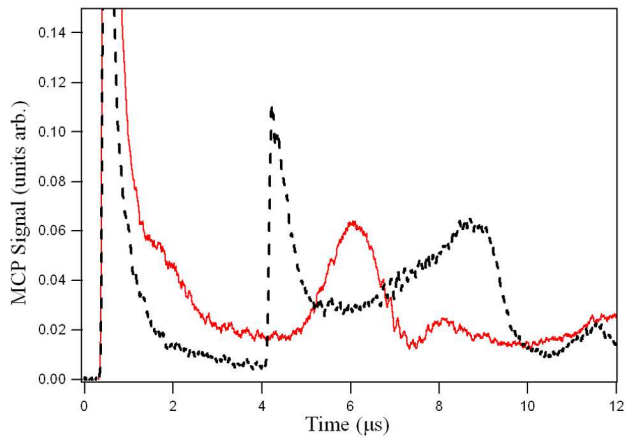}
\caption{(Color Online) A sample of the response to a delayed application of the RF field to the UCP at an applied frequency of 20 MHz ($\Delta$E/k$_B$ = 100 K, 5$\times$10$^5$ total ions, 6 V/m p-p applied RF Field).  The solid, red curve shows the continuous application of the RF field throughout the formation of the UCP, with the resulting electron signal owing to the application of an RF field.  The dashed, black curve shows the RF field being turned on at $\sim$4 $\mu$s and left on continuously afterwards.  Not only does the UCP have a sharp response to the initial application of the RF field, but the location of the resonant response in time can be seen to be shifted later.  The sharp initial response occurs for all UCP conditions in our system.}
\end{figure}

When we delay the application of RF to the UCP until after the initial UCP electrons escape upon formation, we observed a fast ($\leq$175 ns from the onset of the pulse) initial response as seen in Fig. 2.  In Fig. 2, one can see that the response from the delayed RF application produces two peaks, an initial response and a second peak several microseconds later.  The initial peak changes in height as a function of frequency and shows a maximum at a particular frequency depending on the time of the delay.  The second peak is associated with the time of the resonant response to a continuous application of the RF.  The time of this secondary response depends in a complicated way on the time of the RF application, the amplitude of the applied RF field, the UCP expansion rate, the electron temperature, and the frequency of the applied RF compared to the resonant frequency.

Since the response to the applied RF is fast, we were able to excite the plasma using only two cycles of the applied RF field.  The response to a 2-cycle RF pulse is seen in Fig. 3.  The applied RF for our data was always applied with the same initial phase at the onset of the pulse.  In order to determine the number of electrons that escape in response to the application of the RF pulse, we took the difference between the UCP response with and without the RF present (Fig. 3, inset) and integrate.  The integration window is typically 1 $\mu$s after the initial application of the pulse.  If we scan the frequency of our 2-cycle RF at the same time in the plasma evolution, our integrated response shows a broad peak at a particular frequency (Fig. 4).  This peak frequency scales appropriately with the density as we let the UCP expand.  Applying more than 2 cycles of RF to the UCP produces a similar response that is somewhat narrower, but not as narrow as expected based purely on the Fourier Transform of the applied RF.  In Fig. 4, the initial application of the RF is at 12.38 $\mu$s, producing a peak in the electron escape at 12.61 $\mu$s.  This delay between the application of the RF and the electron escape is typical.  For frequencies greater than 10 MHz, the peak of the electron escape is after the 2-cycle pulse is completed.  This means that the response that we see is not because the application of the RF field lowers barrier of the space charge potential at the time the electrons escape \cite{LowFreq}.  

\begin{figure}
\includegraphics{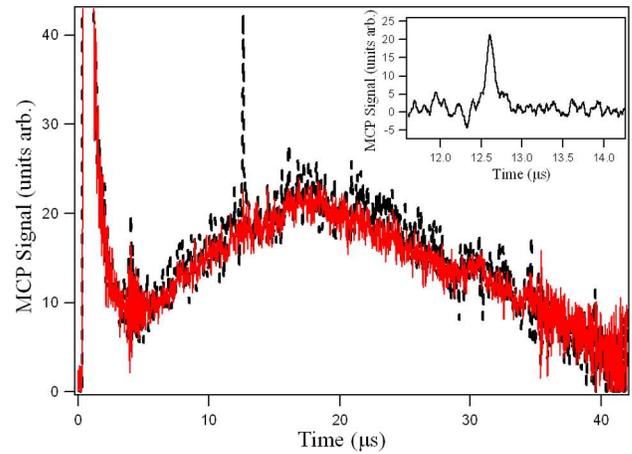}
\caption{(Color Online) A comparison of typical electron escape (red, solid) during UCP expansion to that with an applied 2-cycle RF field (black, dashed). ($\Delta$E/k$_B$ = 100 K, 6.7$\times$10$^5$ total ions, 5 V/m p-p applied RF field at 20 MHz).  The figure is scaled to better see the electron escape after the initial prompt peak.  Inset is the difference between the two signals at the time of the application of the RF.  The signal seen at $\sim$5 $\mu$s is electronic, and not from the UCP.}
\end{figure}

\begin{figure}
\includegraphics{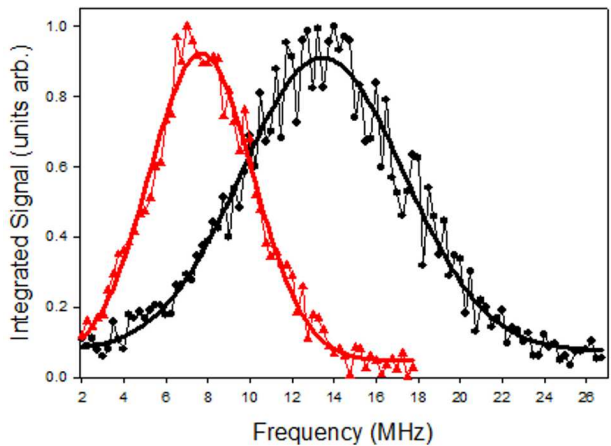}
\caption{(Color Online) The integrated response as a function of applied RF frequency for two different times in the plasma evolution ($\Delta$E/k$_B$ = 10 K, 4.4$\times$10$^5$ total ions, 5 V/m p-p applied RF field).  The black circles are at 39 $\mu$s ($\delta$ = 0.5) in the UCP evolution and the red triangles are at 79 $\mu$s ($\delta$ = 0.9).  This data is consistent with the proper scaling of frequency with the UCP density.}
\end{figure}

In order to interpret our data, we developed a simple model in which we assume that the electron cloud moves as a whole in response to an applied external field within a Gaussian ion density distribution \cite{Essen2005}.  While this would be a poor model under many cycles of applied RF because the non-solid nature of the electron cloud will cause the cloud to de-phase, for 2-cycles it is more applicable.  To determine the resonant response within this model, the electron cloud was displaced from the center of the ion distribution and the restoring force was calculated.  This enabled us to determine the resonant frequency, and its relationship to the peak density of the UCP.  In this model we assume that the peak UCP electron escape occurs when the electron component of the UCP oscillates at its maximum amplitude.

From this model, we were able to calculate the expected ratio of the resonant response frequency to that of the frequency associated with the peak plasma density $\omega/\omega_{peak}$, where $\omega_{peak} = \sqrt{e^2n_{peak}/m_e\epsilon_0}$.  We did this in general for a spherically symmetric, $T$=0 ion/electron distributions by calculating potentials for an $l$=1 perturbation of the electron distribution.  Details of this calculation are given in \cite{SupMaterial}, and the results are seen in Fig. 5.  We compared this infinitesimal displacement analytic calculation to a numerical calculation of the restoring force with a finite displacement in which the electron cloud is displaced by up to $\sim$200 $\mu$m from the center of the ion distribution.  The distance was chosen to correspond to the maximum predicted offset caused by our RF driving fields.  Non-linearity of the restoring force as a function of separation in the numerical model is small, so the two methods for calculating $\omega/\omega_{peak}$ agree to within 2$\%$ over the full range of $\delta$ as seen in Fig. 5.  This is consistent with the sub-3$\%$ frequency shift as a function of applied RF power that is discussed below \cite{ImbalanceCorrection}.

Our simple model predicts that the resonant frequency shifts to higher frequency for a greater charge imbalance.  This can be understood qualitatively as the electrons effectively seeing a larger average ion density as they are more concentrated in the center of the plasma.  To test these calculations experimentally, we performed measurements of the UCP density by using our 2-cycle RF technique while altering the charge imbalance with 1.5 $\mu$s square electric field pulses (up to $\sim$3 V/m).  For initial ionization energies of $\Delta$E/k$_B$ = 10-400 K and initial $\delta$ = 0.15-0.65, we measured the resonant frequency at the same point in the UCP evolution with and without the applied electric field pulse, 1 $\mu$s after the back edge of the pulse.  The pulse typically extracted enough electrons to increase $\delta$ by 0.2.  Since the time is short between the application of the pulse and the measurement of the resonant frequency, we expect that the peak density should be the same in the two cases with and without the pulse.  This produces a frequency shift with the charge imbalance, $\delta$, that can be predicted by our model.  By applying the charge imbalance correction from our model to the measured frequencies, we saw good matching of the values of the extracted peak densities with and without the charge extraction pulse to within 2$\%$, particularly for higher ionization energies ($\geq$ 100 K).  Without applying the correction for the charge imbalance, the extracted values of the peak densities will not match by as much as 25$\%$.  However, this matching of the peak densities seems to break down for the earliest parts of the plasma evolution when the initial ionization energy is low ($\Delta$E/k$_B\sim$10 K), as the frequency shifts are too much to be accounted for by our charge imbalance corrections by $\sim$5$\%$.

For most conditions of our UCPs, we do not have an independent determination of the UCP density in order to directly measure the absolute value of $\omega/\omega_{peak}$.  However, for early enough times in the UCP evolution at low $\Delta$E/k$_B$ (we chose 10 K for this measurement), the expansion rate of the UCP is low, so we can use absorption imaging of our atom cloud to determine the initial density distribution of the UCP without having to model the expansion of the UCP.  By applying an ionization fraction correction to the density distribution from our images, we were able to measure $\omega/\omega_{peak}$ = 0.30 $\pm$ 0.06 for $\delta$ = 17$\%$ and an initial electron energy of $\Delta$E/k$_B$ = 10 K at 5 $\mu$s in the UCP evolution, which corresponds to $\omega/\omega_{peak}$ of 0.37 in our simple model.  This value comes in slightly lower than the predicted value from our model.  Corrections for the asymmetry of the UCP at early times \cite{ImbalanceCorrection} provide somewhat better agreement between the model and the data, but the correction is less than the systematic error of our data.  The general degree of agreement of all of our measurements to the model indicates that a collective electron motion is a reasonable explanation for our observations.

In addition to calculating the resonant frequency for the collective displacement of the electrons in a UCP, we also numerically evaluated whether the fields produced from the motion of the electrons could lead to the increase in the electron escape signal that we observed.  Ideally this calculation would consist of a simulation of all of the electrons' motion, including their interactions.  Such a simulation is beyond the scope of this work.  Instead, we sampled individual electrons' motions as they were influenced by the applied RF fields and the induced electric fields from the overall collective motion of the electrons in the UCP.  For instance, we can start with a $T$ = 0 Gaussian electron/ion distribution with 5$\times$10$^{5}$ ions, an rms radius of 1 mm, and charge imbalances of $\delta$=0.2-0.35.  The electron cloud then oscillates as a whole at an amplitude (typically $\sim$100 $\mu$m) based on our applied external AC field.  The oscillation damps with a time constant chosen in a range of 50-200 ns, which is consistent with the observed damping time of electron motion described below and shown in Fig. 6.  We then place a sample electron with a chosen kinetic energy in various trajectories throughout the plasma distribution.  For our calculations, we sampled electrons with kinetic energies that ranged from 100-350 K$\cdot$k$_B$, and potential depths 400-750 K$\cdot$k$_B$ of energy for this charge distribution.  These calculations show that increase in kinetic energy associated with the motion of the collectively oscillating electron component and acceleration owing to the induced internal fields enhanced by the resonant oscillation can result in individual electrons gaining enough energy to escape from the UCP confinement.  This increase in energy occurs in just a few hundred ns without any binary electron-electron or electron-ion collision processes.  The escape probability of the sample electrons is a function of the initial kinetic energy of the electron, the relative phase of the electron motion to the induced electric fields, as well as the damping time and amplitude of the collective oscillation.  As the damping time is decreased from 200 ns to 50 ns in our calculations, the escape of a sample electron can require up to 6 times the driving amplitude of oscillation.  This decrease in the probability of escape is consistent with observations of the electron response to 2-cycles of RF as discussed below.  While net 90 degree binary collisions (both electron-electron and electron-ion) will limit the total energy transfer for lower energy electrons to be less than what is predicted by our model, the 90 degree collision times for higher energy electrons are long enough \cite{Spitzer1962} (10 times or more greater than the predicted electron transit time across the UCP) for this energy transfer calculation to be reliable.

\begin{figure}
\includegraphics{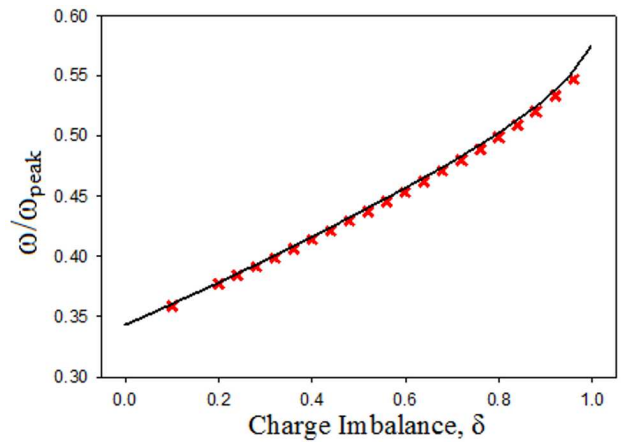}
\caption{(Color Online) A plot of the ratio of the measured resonant frequency to the frequency associated with the peak density in the UCP as given by $\omega_{peak} = \sqrt{e^2n_{peak}/m_e\epsilon_0}$.  The black line shows the result of the average density calculation discussed in \cite{SupMaterial}.  The red X's show the results from a numerical simulation in which the electron and ion distributions are offset from each other, and a restoring force is calculated.}
\end{figure}

\begin{figure}
\includegraphics{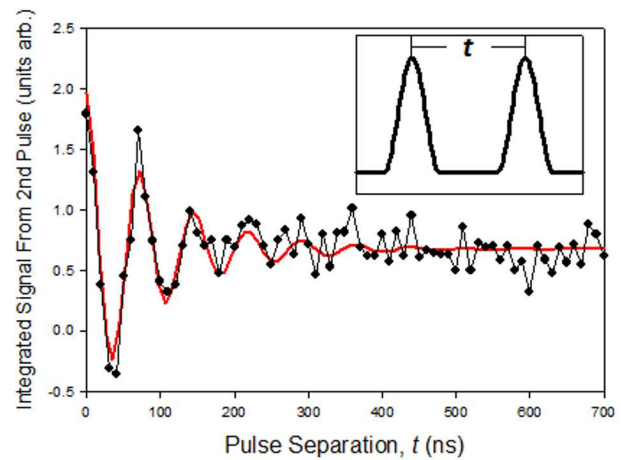}
\caption{(Color Online) An example of the response of the plasma to 2 electric field pulses (FWHM of 12.5 ns per pulse) at $\Delta$E/k$_B$ = 10 K (55 $\mu$s in the plasma evolution).  The integrated response here (black circles) is the response to the second pulse specifically.  The red curve shows a fit of an exponentially decaying, oscillating function to the data.  This fit has a frequency of 13.67 $\pm$ 0.23 MHz as compared to a resonant frequency measurement of 13.31 $\pm$ 0.12 MHz in this case.  The decay constant of our fit in this case is 103 $\pm$ 14 ns.  The inset is an example of the set of the pulses.  The pulses are identical, with their peaks separated by a time $t$.}
\end{figure}

In modeling the UCP response as a collective motion of the electrons, we should expect to be able to excite this motion by applying sharp electric field pulses to the UCP.  By applying an initial electric field pulse (FWHM of 12.5 ns), we can excite motion of the electrons along the axial direction.  After a time delay, we can apply a second pulse and measure the number of electrons that escape the plasma.  We observe clear oscillating behavior in the electron response as seen in Fig. 6.  The frequency of oscillation corresponds to the peak frequency of a 2-cycle RF sweep measurement at the same time in the UCP evolution.  This frequency follows that of the resonant frequency at different points in the UCP expansion, and thus scales with the density.  The oscillating signal decays in time.  That decay could be due to de-phasing of the electron cloud during oscillations or damping of the oscillations through collisions.  At low initial ionization energy ($\Delta$E/k$_B$ $\sim$10 K) and early in the plasma evolution, applying these sharp electric field pulses to the UCP can show an over-damped behavior in the electron response to the pulses in which no oscillations are present.  Under these conditions, the UCP collision rates are much higher than at other points in the UCP evolution, and a much higher damping of the free oscillations from binary collisions is expected to occur.  This over-damped behavior under these UCP conditions is a possible reason why our predicted charge imbalance corrections from our model begin to fail, and will be the subject of further investigations.

We predict, based on our observations, that the energy transfer mechanism that is described in this article will not be as prevalent in other UCP experiments where the peak densities are 1 to 2 orders of magnitude higher \cite{Kulin2000,Zhang2008,Fletcher2006}.  The restoring force from the ion distribution increases with density, which will decrease the magnitude of the electron cloud oscillation.  We were able to measure the effect of the density on the UCP response in our system.  We measured the resonant frequency at many different times in the UCP evolution using our 2-cycle RF burst technique.  By scaling the electron signal produced by the number of electrons remaining in the UCP, we observed that the effective response size to the 2-cycle RF decreases greatly as the density increases.  If we try to compare the results presented here to the theoretical predictions of \cite{Lyubonko2012}, we find that the results do not match the theory.  However, the experimental conditions for our system, namely the magnitude of the applied RF, violate the assumption of small density perturbations in \cite{Lyubonko2012}.  Therefore, it is not surprising that the theory in \cite{Lyubonko2012} is not applicable to the data presented here.

To further illustrate the difference of our response mechanism to those in other UCP systems, we can compare the response of the UCP to continuous RF and 2-cycles of RF by using the measured resonant response from the two techniques to calculate the RMS size, $\sigma$, as the UCP evolves.  For the continuous RF case, we applied a single RF frequency throughout the expansion of the UCP.  We observed that by changing the amplitude of the applied RF, the time of this additional electron escape can shift significantly for our experimental conditions as seen in Fig. 7a, particularly at low initial ionization energies.  This makes the precise determination of the time at which the resonant response conditions would occur in the absence of an RF field problematic.  At comparatively high initial ionization energy ($>$100 K), the time of the resonance as a function of amplitude could be extrapolated back to zero power so that we could determine the time that the resonance conditions were met when the UCP expanded in the absence of applied RF.  We interpret this shift with applied amplitude as being due to continuous application of RF affecting the expansion of the UCP.  This effect can be observed by measuring the effect of the RF on the UCP lifetime, which is defined as the time it takes for the last of the electron signal to reach the detector.  The lifetime of the UCP is greatly shortened for lower initial electron energies as can be seen in the inset figure in Fig. 7a.

\begin{figure}
\includegraphics{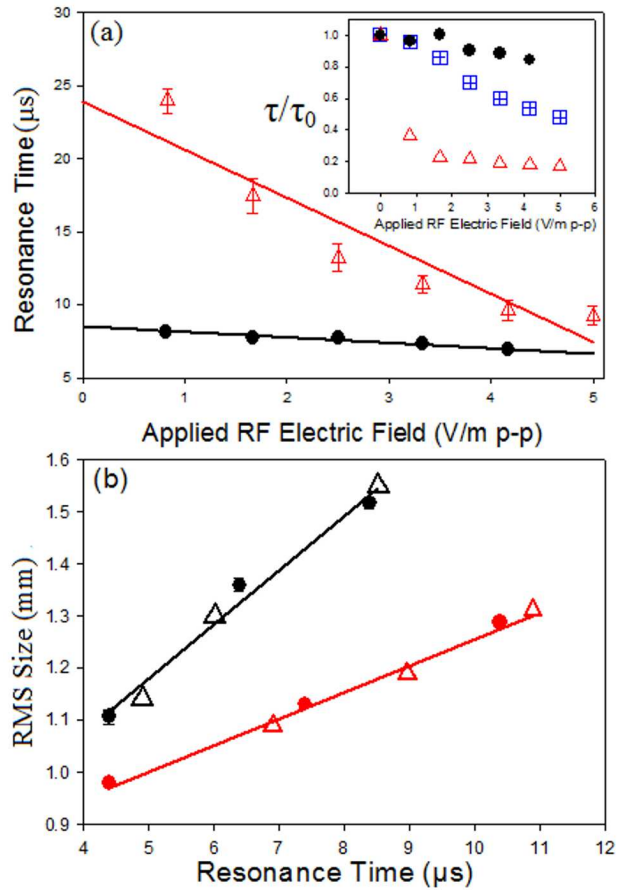}
\caption{(Color Online) (a) An example of how the resonance peak time for continuous RF shifts with the applied RF voltage.  At $\Delta$E/k$_B$ = 10 K (red triangles, open), one can see that the extrapolation is not linear, and the zero power peak time is not determined by a linear fit (12 MHz applied RF).  At higher initial ionization energies (closed black circles here are at $\Delta$E/k$_B$ = 400 K) the extrapolation is linear and the zero power peak time can be determined (16 MHz applied RF).  The inset figure shows the ratio of the lifetime of the UCP with an applied RF field to the lifetime with out the field present ($\tau_0$ = 25, 50, and 110 $\mu$s for $\Delta$E/k$_B$ = 400, 100, and 10 K respectively).    In (b), we compare that extrapolation for a few different frequencies of continuous RF (triangles, open) with the measured frequencies at specific times with the 2-Cycle RF method (circles, closed).  In this data, for $\Delta$E/k$_B$ = 100 K (red) and 400 K (black), these data show a consistent trend in time. The applied RF voltage is from point A in Fig. 1.}
\end{figure}

We can extrapolate multiple frequencies of continuous RF in this way and plot them with the results of 2-cycle RF sweeps at multiple times as seen in Fig. 7b (Note: Changing the amplitude of the 2-cycle RF sweeps by a factor of 3 changes the peak frequency response by only ~3$\%$).  From the measured frequencies and charge imbalance, $\delta$, we can use our model to calculate the peak density, $n_{peak}$, and RMS size, $\sigma$ of the UCP.  We can calculate $\sigma$ from a spherical Gaussian distribution to be $\sigma=(N_{ion}/(2\pi)^{3/2}n_{peak})^{1/3}$, where $N_{ion}$ is the total number of ions and electrons.  We observed that the RMS size, $\sigma$, as a function time shows no significant (i.e. greater than ~3$\%$) difference between the two different techniques.  This degree of agreement shows that the mechanism for the resonant response is likely the same in both cases.  This further confirms that our experimental parameters put us in a regime where the theory from \cite{Lyubonko2012} does not apply, because the charge imbalance correction to $n_{peak}$ would be very different from our model, particularly at early times in the UCP evolution as presented here.

In summary, we observed that 2-cycle RF pulses resulted in a rapid, density-dependent response from ultracold plasmas.  Our observations are consistent with a collective electron motion playing a strong role in redistributing the energy in the UCP in response to an external perturbation.  Understanding that such a mechanism is present will be important in properly interpreting the results of any experiment where the UCP is subjected to rapid external perturbations.  We developed a simple model in which we treat the electron cloud as moving as a whole to explore the general features of the physics that we think is responsible for the rapid electron response.  This model showed that significant energy can be transferred to the escaping electrons quickly without the need for binary collisions.  In addition to exploring the UCP response to RF fields, 2-cycle RF pulses provide an additional way to measure the UCP density at specific times and thus expansion rate with less amplitude sensitivity than applying continuous RF.

This work was supported by the Monfort Foundation and by the Air Force Office of Scientific Research, grant number FA9550-08-1-0031.  The authors also acknowledge the help of Matthew Heine in the construction of our experimental apparatus.


\begin{thebibliography}{15}
\bibitem{LowFreq} At low frequencies (relative to the peak frequency) under certain UCP conditions, the effect of the applied RF field on the confining potential can be observed.  However, this typically only occurs at high driving amplitudes of the applied RF and can be easily avoided.  The response in these situations has different characteristic behavior, and does not have any effect our peak frequency determination for our data.
\bibitem{ImbalanceCorrection} The UCPs for our system have an initial charge distribution that is not precisely spherically symmetric as a consequence of the transfer of the atoms in a magnetic trap.  The corrections to $\omega/\omega_{peak}$ owing to this were calculated to be less than 10$\%$ (shifting $\omega/\omega_{peak}$ to lower values) over the full range of $\delta$.
\end{thebibliography}
\end{document}